# The early history of Marine Cloud Brightening (MCB)

## The legacy of John Latham and Stephen Salter

*Alan Gadian[1] - and a host of others.*
[1] *National Centre for Atmospheric Sciences and the University of Leeds, UK.*

Correspondence emails: alan.gadian@ncas.ac.uk and a.m.gadian@leeds.ac.uk

**Orchid id:** 0000-0001-9890-403X

**Date:** October 2025

**Abstract**

This paper discusses the initial development of **Marine Cloud Brightening (MCB)** as a theoretical idea, from its inception as a cloud microphysics process in ~1990 to the full-blown concept by 2015. It primarily focuses on the work of founders John Latham and Stephen Salter and their contributions. Recently the concept has been developed further, e.g. in the UK ARIA project, as a prospective method to ameliorate the Earth's rapid warming.

### 1. Introduction.

Our purpose is to shed some light on the history and early research-and-development work of what is now called "**Marine Cloud Brightening" (MCB)**, from the initial work by Latham until its more general acceptance by 2015, as a method to cool the planet. At its outset it was called "Marine Cloud Whitening" (Gadian et al. 2009), but. soon after, Latham changed the name to "Marine Cloud Brightening". This terminology better encapsulated the aim of 'brightening' the planet by increasing Earth's albedo to reflect more incoming solar radiation back to space. Incidentally, some now prefer the term 'rebrightening' (https://rebrighten.org).

Latham's central idea was to use sea-salt of a defined optimal size to spray into marine stratocumulus-layer clouds to increase the number of cloud-droplets in these clouds at the expense of slightly reducing the mean droplet radius. This has the effect of increasing the reflectance (albedo) of these clouds to incoming solar radiation. Salter liked to illustrate this dramatic effect with the help of two glass jars containing glass beads, as shown in Figure 1. The beads of smaller diameter appear brighter/whiter than the larger beads, in visible light.

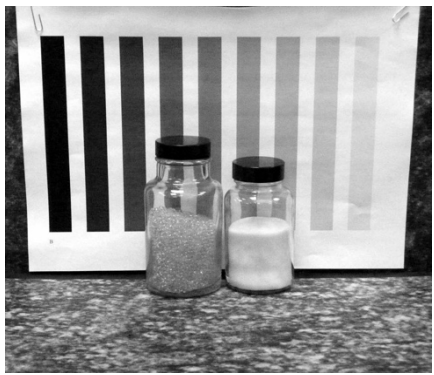

**Figure 1.** The left-hand jar is filled with glass beads 4mm in size and has an albedo of ~0.6. The right-hand jar contains glass beads of size ~40 microns and has a much higher albedo of ~0.9. Both jars contain the same volume of glass beads. (*courtesy Stephen Salter*)

## 2. Planet Earth is rapidly warming.

Over the past decades there have been numerous urgent warnings of rapid warming. The Charney report (1979) declared that for a double $CO_2$ atmosphere, warming would raise the Earth's Equilibrium temperature by ~4.5C. He produced this value using standard radiation transfer theory without using any climate models. The basic arguments were presented at the 1978 EGU conference but largely ignored. Lovelock (2006) in 'The Revenge of Gaia' stated that within 30 years the Earth would suffer increasing weather catastrophes. Latham (2002, 2008) likewise warned of the effects of doubling $CO_2$ in the atmosphere, which he stated would increase IR (Infrared Radiation) 'radiative forcing' by ~3.7 W/m$^2$. (Ramaswamy et. al. 2001).

There are many publications now appearing which highlight the magnitude of the problem. To cite but three: the earth's energy imbalance has more than doubled in recent decades (Mauritsen et. al., 2025), being measured now as ~1.5 W/m$^2$. Tselioudis et al. (2025) produces a warming rate of 0.37 W/m$^2$ per decade. Yuan et al. (2025) provides a figure of 0.33 W/m$^2$ per decade. The consequences are now becoming very apparent. Quoting from the conclusion of Drijfhout *et al* (2025) on the shutdown of the northern Atlantic Overturning Circulation, "*Of particular concern is our finding that deep convection in many models stops in the next decade or two, and that this is a tipping point which pushes the northern AMOC into a terminal decline from which it will take centuries to recover, if at all*."

## 3. Founders of Marine Cloud Brightening.

John Latham was an internationally renowned atmospheric scientist who after studying at Imperial College went to work at UMIST and NCAR. He was chair of the International Cloud Physics and Atmospheric Electricity conferences. He was an internationally renowned poet and some poets argue he could have become poet laureate if he had chosen that route. By the late 1990's he realised the planet was suffering from rapid greenhouse warming and he promoted the idea of cooling the planet by increasing the reflectance of clouds, in doing so inventing "Marine Cloud Brightening/Whitening". Some of his publications on MCB include Latham (2008, 2010, 2012, 2012b, 2013, 2014) and Bower et. al. (2006). He appreciated that water-vapour feedback – see science section below – was important, and that increasing planetary albedo was the best temporary fix. His obituary can be found in the Guardian (Gadian, 2021)

Stephen Salter was an internationally famous engineer. He had vast experience as a fluid mechanics engineer, based in the University of Edinburgh. He invented the "nodding duck" and various other mechanical devices including a hydraulic system for the UK Americas cup boat. He joined John Latham in early 2000's and produced engineering salt spray equipment. He liaised with Armand Neukermans (Neukermans et al., 2014) and produced many scientific conference and science publications, (Salter et al. 2008 & 2013), as well as contributing to Latham's publications. He died in 2024 (Guardian obituary, Brown, (2024)). His practical legacy was to set up the Lothian School of Technology, Edinburgh, where spray systems are now being constructed.

John Latham and Steven Salter worked together on both theory and engineering design, and enabled the further development of the technology by involving and enthusing other scientists. From the outset, Alan Gadian was fully involved in the project. He supervised two PhD students - Laura Kettles and Ben Parkes - who conducted the global modelling simulations, and Mirek Andrejcuzk, who completed the high-resolution cloud simulations.

## 4. Foundational principles of Marine Cloud Brightening.

The central physics behind MCB is based on the work of Twomey (1977, 1991). He was the first to highlight the influence of CCN (Cloud Condensation Nuclei) – i.e. polluting aerosols - on the short-wave albedo of low-level clouds. That 'influence' became known as the 'Twomey Effect'. On balance, clouds produce a cooling effect, corresponding to a globally averaged negative net forcing of approximately ~13 W m$^2$ (Ramanathan et al. 1989). Low-level cloud marine stratiform clouds covering about a quarter of the oceanic surface (Charlson et al. 1987) - with albedos in the range 0.3–0.7 (Schwartz & Slingo 1996) - can provide much of this forcing.

Slingo (1990, 1982) described the sensitivity of the earth's radiation budget to changes in low level clouds. This work was rapidly followed by Latham's (1990). It argued that modification of the albedo of low-level clouds could be used to control global warning. Later this was termed as the first *Indirect Effect*. Additionally, Albrecht (1989) suggested that increases in aerosol concentrations over oceans may increase the amount of low-level cloudiness. This further increases the earth's albedo and cooling of the earth's surface: the second *Indirect Effect*.

Recent results from Chen et al. (2024), suggest that thanks to the combination of these two *Effects*, albedo-increase will be larger than was originally thought likely. They find that there is a greater sensitivity of climate to radiative forcings. The mitigation of global warming (MCB) is possible and effective in humid and stable conditions in the sub-tropics with strong incoming solar radiation. This work and these results directly support Latham's (1990) proposal.

'Marine Cloud Whitening' is described in layman terms in (Gadian et al., 2009). Soon after the publication of that paper, the process was renamed 'Marine Cloud Brightening'.

## 4.1 The Politics of MCB

From the outset there was much polarisation within the research community, and resistance to anything describable as 'geo-engineering". It was extremely difficult to get the important paper Latham (2002) published at all, and it was opposed, ironically, by some of those scientists and organisations who now participate in research projects in ARIA - a UK funding body focused largely on MCB and comparable 'geo-engineering' concepts.

The Royal Society report, Sheppard (2009) – further referred to below – substantially miscalculated the cost of MCB. The overseeing committee for that report, despite Latham and Salter's best efforts, would not publicly make the correction; it is not unreasonable to speculate that their preference for Stratospheric Sulphur Injection as a potential climate-cooling technology, may have been a factor in this significant oversight.

Some senior scientists and organisations, including the UK Research Councils NERC and EPSRC, precluded applications to research into the science of MCB. Senior scientists at certain universities furthered this systematic neglect, despite knowing little about cloud dynamics and microphysics. Nonetheless, MCB was strongly supported by Paul Crutzen (Nobel prize for SAI), James Lovelock (GAIA), Armand Neukermans (USA) and John Pyle (head of Chemistry at Cambridge), among many others. But even the presence of a large body of experts failed to overcome the negativity created by the 'anti-geoengineering' lobby or the rigidity of the SAI lobby (who were all-too-little concerned about SAI's adverse side effects).

In MCB, there remained questions, for Latham, Salter and Gadian, about optimal cloud droplet size, including the effects of turbulent droplet broadening and dynamics, which needed further consideration. The possible unintended and consequential impacts on precipitation, for example, would call for further study. Salter and Gadian requested research into these important impacts using the 'coded modulation' technique (invented by David Mackay, a UK government chief scientist). Recently funded UK ARIA projects do not address these issues.

Possibly, Chinese scientists may follow up these ideas along with the other unresolved questions stemming from MCB. Stephen Salter's laboratories have created technology which could produce spray of the correct size. Other groups in Australia (see rebrighten.org) are working on this technology too. Salter's and Neukermans' groups were the first to produce the correct size spray in ~ 2009. But ARIA has opted not to follow up their spray-size research. And their deployment of non-disclosure and IPR agreements is something that Latham, Salter and Gadian were always opposed to, on the grounds that science should be open to all.

Other scientists currently argue that '***due diligence***' has not been followed in MCB science. There is a lack of basic scientific evidence and understanding of the principles underlying

MCB. Indeed, resistance to doing basic research into MCB is apparent. This is exemplified in the October 2025 Senate hearing on (objections to) emissions of chemicals into the atmosphere. This hearing does not even consider ARIA's plans to conduct experiments. In the harsh light of reality, though: if Drijfhout *et al* (2025) predictions on the breakdown of the AMOC are correct, then there may be rapidly growing pressure to do something soon.

## 5. The science and initial developments.

## 5.1 The initial publications on MCB.

Latham (1990, 2002) proposed a possible technique for ameliorating global warming by controlled albedo enhancement. The unseeded original cloud-droplet number-concentration ($N_0$) is increased to a new seeded concentration ($N$) in maritime stratocumulus clouds. There is a corresponding increase ($\delta A$) in their albedo. This is called the '*first indirect effect*', or 'Twomey effect' (1977). In their longevity, these *effects* are called the '*second indirect effect*', or 'Albrecht effect'- Albrecht (1989). Schwartz & Slingo (1996) argued that the cloud albedo increase resulting from seeding the clouds from $N_0$ to $N$ with sea-salt cloud condensation nuclei (CCN) would produce an albedo change such that $\delta A = 0.75*\ln(N/N_0)$

Latham (1990, 2002) calculated that for an ideal(?) droplet diameter of ~0.8 $\mu$m and associated salt mass of $10^{-17}$ kg, the total (global) seawater volumetric dissemination-rate required, to produce the required doubling of N in all suitable marine stratocumulus clouds is approximately 30 $m^3 s^{-1}$. The distribution should be monodispersed, to avoid the production of large and ultra-giant nuclei. Mono-dispersity will make the clouds more stable, inhibiting coalescence and thus associated drizzle formation.

In the Latham et al (2008) paper is the fundamental description of Marine Cloud Brightening. In it the microphysical principles, a first assessment, the theoretical and the initial atmosphere-only climate-modelling results are discussed. Choosing three most persistent maritime stratocumulus regions, representing ~5% of the surface, could provide an albedo response which could balance an IR warming of ~3.7 $W/m^2$. This figure of ~3.7 $W/m^2$ is the suggested warming generated by ~double $CO_2$ radiative forcing (Ramanathan et al. (1989)). The work represented a collaboration between UK (Leeds, Manchester and Edinburgh) and MMM and Climate divisions at NCAR. It is of course not possible to summarise this paper in a paragraph. It needs to be read and understood in detail, to understand the principles of MCB.

The corresponding technology paper by Salter et al. (2008) proposed self-propelled ships which could be utilised to provide suitable platforms for the spray distribution. Salter calculated that just 2000 vessels are required, which could be moved appropriately to produce optimal albedo-modification – although, again, the focus would be on the three selected regions.

The two papers very much complement each other. At this time, Salter and Neukermans (2014) had produced the only two spray emission-devices. The Salter / Edinburgh / Ocean Cooling Technology groups have continued to develop the spray devices.

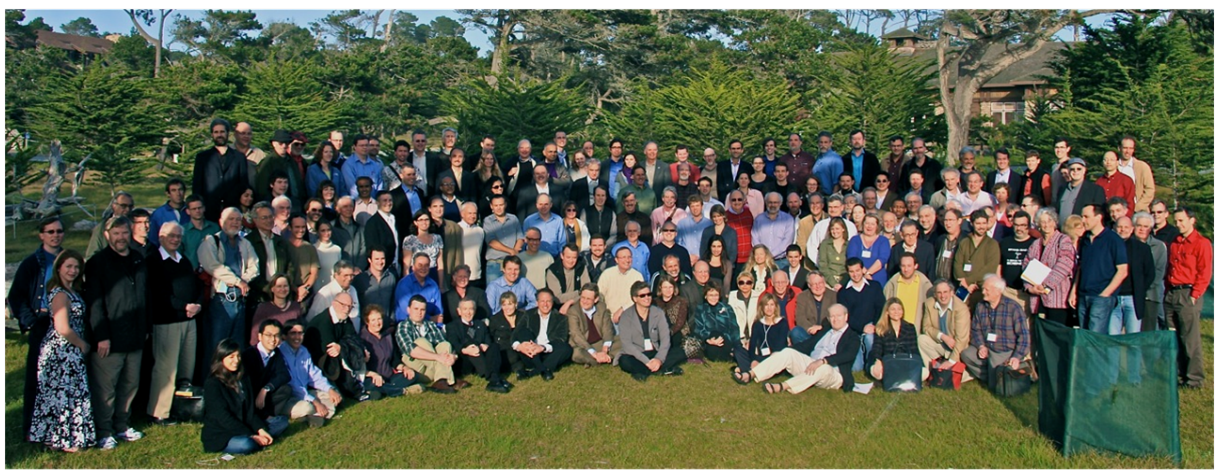

**Figure 2**: Asilomar conference 2009. On geoengineering, 2009, including Stephen Salter, (left) John Latham (4[th] from left), Paul Crutzen (centre). Group Photograph

( https://web.archive.org/web/20130831080727/http://climateresponsefund.org/images/Conference/finalfinalreport.pdf )

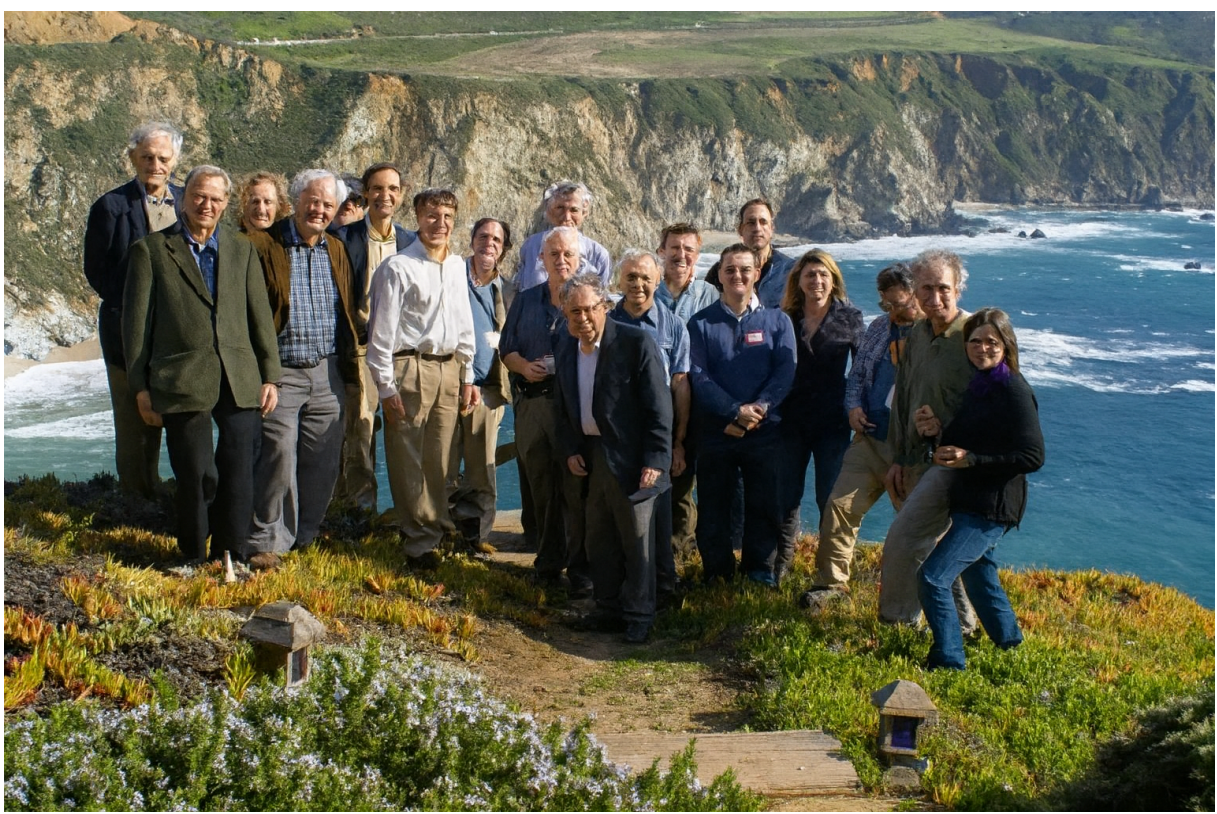

**Figure 3.** Big Sur, meeting after the Asilomar Conference 2009. including S. Salter (left), John Latham (4[th] from left), A. Neukermans, A. Gadian (centre back), P. Crutzen (centre), R. Wood (5[th] from the right), P. Rasch (2[nd] from the right) (courtesy Armand Neukermans)

### 5.2 Publications from other scientists related to the science of MCB.

Several other important publications followed these two papers. The Royal Society report, Sheppard (2009), summarised the then-current Solar Radiation Management techniques. Unfortunately, the costing of MCB was overestimated by an order of magnitude, suggesting it was not financially viable. Jones et al. (2009, 2011 & 2012), based at the UK Met Office, further developed the modelling responses using climate models. They compared the relative efficacy of SAI (Stratospheric Sulphur Injection) as in Crutzen (2006), with MCB in an Earth system model. Crutzen later went on to argue that he thought MCB was safer and preferable.

It is difficult to assess the relative contributions from just one model simulation. Rasch et al. (2009), used the NCAR CCSM model to assess the MCB influence on sea ice and climate system. Bala et al. (2010) discussed the effects of albedo enhancement of marine clouds using MCB. He detailed the impacts on the hydrological cycle, again using the NCAR climate model. Using high resolution cloud models, Stevens and Feingold (2009) analysed and "untangled" the aerosol effects on clouds and precipitation in such a cloud system.

Wang and Feingold (2009, a,b) analysed stratocumulus cloud decks and the impact of ship tracks, examining their dynamics and microphysical properties. These papers followed on from and used the data from the Vocals / Vamos experiment, which studied the properties of stratocumulus clouds, (Wood et al. 2007). All contributed to the discussion of the efficacy of MCB. Finally, Kravitz et al. (2011) lead the Geoengineering Model Intercomparison Project (GeoMIP). This was the first time a multi-model intercomparison project had been considered with MCB and other geoengineering ideas.

## 5.3 Associated meetings and presentations on MCB

A significant event was the Asilomar meeting (2009), where over 200 scientists met and discussed Geo-Engineering, Solar Radiation Management and Carbon-Sequestration techniques. This meeting was based on a previously very successful meeting on DNA advances and how that technology could be managed.

Unfortunately, the topic of regulating the approach to climate intervention strategies proved more difficult. Figure 2 shows the group photograph of the conference. Immediately afterwards, those scientists interested in the development of Marine Cloud Brightening met at Big Sur, to discuss how to develop these ideas. This provided a useful start to the collaboration (Figure 3). It was also followed by a conference meeting for over 300 scientists (PACC conference) in Moscow in 2011 convened by Yuri Izrael. Latham's and Salter's MCB ideas were presented by Gadian and Salter, alongside other studies, e.g. low-level SAI experiments being conducted in Russia. Presentations of MCB were made by Latham, Salter and Gadian at the Royal Society in London, National Academy of Sciences Washington, and at other conferences.

Some funding indirectly from the Bill Gates foundation had been obtained to develop SRM techniques. Most funds were allocated to the development of SAI and "Carbon Dioxide Removal" (CDR) techniques, but a small sum was made available to develop MCB modelling work in the UK, which lead to modelling projects by Gadian. Parkes and Andrejcuk. Ultimately this led to further publications in 2012 – rounding off initial development work in MCB.

## 6. The final research publications in the initial MCB development phase.

Brian Launder arranged a special edition of the *Philosophical Transactions of the Royal Society* in 2012 to discuss the latest research into Marine Cloud Brightening. Latham et al. (2012) encapsulated the culmination of this research effort. The forty-five-page paper / report was a result of the collaboration of groups at University of Washington (lead by Wood), PNNL (lead by Rasch), Manchester (lead by Latham / Coe / Connolly), Leeds (lead by Gadian) and FICER (lead by Neukermans). The global modelling results used HadGEM1-2, a fully coupled ocean model, in contrast to atmosphere-only results reported in Latham et al. (2008).

The fully coupled model simulations were an improvement but largely replicated the previous atmosphere-only climate model computations. The precipitation changes in the double-$CO_2$ simulation were far more extreme and variable compared with the MCB / double-$CO_2$ simulations, again for the three-region seeding regime. The addition of MCB reduced the average precipitation changes as well as extremes. Simulations using the whole-ocean seeding scenario, produced significant global cooling.

High-resolution Stratocumulus modelling - including examples of ship tracks - was included, and related to the work of Feingold et al. (2010). Discussion by Connolly of parcel model calculations to assist with choice of optimal drop-size was incorporated in the presentation.

Redesign of the spray-devices and of possible spray-platforms was described by Salter. The design of a possible field experiment design by Wood was included, following on from the successful VOCALS Stratocumulus project (Wood 2007). Again, it is not possible to summarise the work presented in Latham et al. (2012); it needs be read in depth to understand the conclusions reached - and outstanding questions raised - by this MCB research. The weather / climate impacts of MCB implementation, other than precipitation-changes mentioned above in Latham et al. (2012), are discussed below in separate publications.

## 6.1 Coral Bleaching

Reduction of bleaching of coral reefs, especially off the Eastern Australian coast, is of great concern in our warming climate. The duration of short temperature-spikes is instrumental in causing the strongest bleaching. Applying the results from the model simulations (Latham et al., 2013) demonstrated that the bleaching effects were greatly reduced, if not reversed, with the application of the three-region seeding.

It was apparent from the results, that the ocean and atmosphere circulations were important in the transfer of cooler waters from the Eastern Pacific cooler zones. The model simulation was set up so that there was a gradual increase in $CO_2$ from (then) present-day values in 2000 to double $CO_2$ levels 2045. After a further spin-up of 25 years at these $CO_2$ levels, the ocean and atmosphere were considered to be in a steady state, having had time to form a modified steady circulation. Data generated for the years 2070 to 2090 was used in the study. The conclusion was that the effects of MCB, even though only applied in three relatively small areas, would impact on the temperatures of the whole global maritime regions.

## 6.2 Hurricane amelioration in the North Atlantic.

MCB would have impact on 'Hurricane Development' in the southern North Atlantic. Latham et al. (2012b) discussed how the use of MCB - again seeding only in the three regions - reduced Sea Surface Temperatures (SSTs) in what was termed the "Hurricane Development Zone", a region between the West coast of North Africa and the Gulf of Mexico. It was concluded that the impact of MCB would be to reduce intensity of *severe category* hurricanes, following the theory enounced in publications on the development and growth of hurricanes. The simulations predicted that SSTs would reduce to levels close to pre-industrial levels.

## 6.3 Regional Applications.

Further analysis of modelling results in Latham et al. (2012) provided insight into the regional effects of MCB. Latham et al (2014) discusses changes to polar ice-sheets - in the Arctic in particular - in extent and volume. According to contributions from Peter Wadhams, the reduction of the polar ice-sheets would cease, and the ice-sheets would ultimately regrow. With whole-ocean seeding, planetary albedo would further increase, the mechanism for this increase being increased aerosol-reflection, sometimes referred to as 'clear air brightening'. The increased planetary albedo would further enable ice-sheets to advance equatorward, to an extent greater even than in pre-industrial times. This radiative solar reduction, even within a double-$CO_2$ atmosphere, would dominate the earth's greenhouse radiative-forcing impact.

## 6.4 Meridional Heat Flux.

A key question is this: what is the mechanism of MCB cooling? And how does the albedo change associated with increasing the *droplet-number* and marginally decreasing *droplet effective radius* in sub-tropical stratocumulus, have such an impact in polar regions?

The meridional polar heat-fluxes are the mechanism by which the planet cools the equator and warms the polar regions. The atmosphere's role is primarily as a heat-engine to transfer heat and energy. Using the data from model simulations (Latham et al. 2012), we can calculate how the meridional heat-flux transfers heat - and just as importantly moisture - from the tropics to the poles, in a double-$CO_2$ three-region-seeding simulation.

Parkes et al. (2012) showed that the poleward heat-fluxes in this scenario were very similar to those of the 'control experiment' of actual present-day conditions at the time. The ice-melt would be stabilised. For the double-$CO_2$ simulation with no MCB, peak meridional heat-flux shows increases from approximately 5.1 petawatts (control) to 5.3 petawatts (double-$CO_2$). This difference is sufficient to produce large polar ice-melt. In all maritime seeding simulation scenarios, maximum polar heat-flux is approximately 4.5 Petawatts, and polar ice-fields grow.

### 6.5 Coded modulation.

It is a significant challenge to try to estimate unintended consequences of implementing MCB.

With regard to precipitation: it seems that when MCB is implemented, extremes of precipitation are substantially reduced by comparison with a zero-seeding double-$CO_2$ scenario. Salter (2013), Salter and Gadian (2013) proposed that the effects of turning the seeding 'on and off' could be simulated using the application of *coded modulation* to the seeding forcing-function.

Parkes (2012) divided the ocean into 89 separate equal areas and applied a *coded modulation* algorithm (Salter and Gadian, 2013) to simulate the effects of turning the seeding 'on and off' using a two-week random forcing-perturbation to the drop-size concentration.

For example, transfer functions were generated to correlate activation of a seeding source with effects on precipitation variances. For impact zones in the North and South Amazon land regions (Salter and Gadian, 2013), turning the seeding 'on and off' in different maritime areas, could produce increases or decreases in precipitation in those specific zones in the Amazon. But the results are very speculative. Changes in precipitation-patterns occurred in different areas. Calculations would have to be repeated using higher-resolution tools. More research needs to be done to address such intentional (or unintentional) consequences of using MCB.

### 6.6 Optimal spray droplet-size.

One significant challenge that is still outstanding, is the determination of the optimal spray droplet-size. Connolly et al. (2014) made a significant contribution to Latham et al. (2012). Their analytical and microphysical models define the factors that determine the most efficient spray droplet-size. They produced an optimal value of 30-100nm (Connolly et al., 2014).

Andrejczuk et al. (2012, 2014) approached the question from a different direction. They used a very high-resolution (20m) fluid-dynamical model, but it incorporated *lagrangian* trajectories of droplet-distributions. This approach avoids the problems of numerical diffusion found in "bin" microphysical cloud models and facilitates creation of multi-spectral droplet-distributions. They found that 500nm wet diameter was an ideal size to use, more in agreement with Latham et al. (2008). But the range and scope of sensitivity were limited. Calculations were only in two dimensions due to big computational costs. The clear advantage was that turbulent and mixing processes were well described. These were not really included in Connolly et al. (2014).

What is the optimal spray droplet-size is still uncertain. To this day it is an important research question. And it would appear that it is not a question that is currently being addressed, despite the many millions of dollars being invested in the development of spray devices.

### 7. Summary

This brief historical and academic resumé describes the initial development of 'Marine Cloud Brightening', led by Latham and Salter. In the final publications, their work was supported by a collaborative group of about twenty scientists. Despite widespread opposition to the very concept of 'geoengineering' and the prevailing focus on 'Stratospheric Sulphur Injection' techniques, interest in the subject remains strong. MCB can be closely controlled - quickly terminated or modified. That confers a large advantage over most geoengineering techniques.

The author apologises to all those scientists who were involved in MCB research in this early period in its history, who are not named here and whose research is not specifically discussed. The overall aim of this paper is simply to provide a record of the initial development of MCB.